\documentclass[times,10pt,twocolumn]{article}

\usepackage{latex8}
\usepackage{times}
\usepackage{latexsym}
\usepackage{amsmath}
\usepackage{amsfonts}
\usepackage[english]{babel}
\usepackage{array}

\newtheorem{definition}{Definition}[section]

\begin{document}

\title{Measuring Similarity in Large-scale Folksonomies}

\author{
Giovanni Quattrone$^1$, Emilio Ferrara$^2$, Pasquale De Meo$^3$, Licia Capra$^1$\\
$^1$\emph{Dept. of Computer Science, University College London, UK}\\%, WC1E 6BT, UK}\\
$^2$\emph{Dept. of Mathematics, University of Messina, IT}\\%, V.le F. Stagno D'Alcontres 31, 98166, IT}\\
$^3$\emph{Dept. of Physics, Informatics Section, University of Messina, IT}\\%%, V.le F. Stagno D'Alcontres 31, 98166, IT}\\
E-mail: \{g.quattrone,l.capra\}@cs.ucl.ac.uk; \{eferrara,pdemeo\}@unime.it
}

\maketitle

\begin{abstract}
Social (or folksonomic) tagging has become a very popular way to describe content within Web 2.0
websites. Unlike taxonomies, which overimpose a hierarchical categorisation of content,
folksonomies  enable end-users to freely create and choose the categories (in this case, tags) that
best describe some content. However, as tags are informally defined, continually changing, and
ungoverned, social tagging has often been criticised for lowering, rather than increasing, the
efficiency of searching, due to the number of synonyms, homonyms, polysemy, as well as the
heterogeneity of users and the noise they introduce. To address this issue, a variety of approaches
have been proposed that recommend users what tags to use, both when labelling and when looking for
resources. 
%Whilst these techniques have proved to increase the accuracy of searches,  we argue that they do so in unrealistic settings, where the datasets used during tests have been manipulated,  typically computing a $p$-core that projects the folksonomy onto a much denser space. 
As we illustrate in this paper, real world folksonomies are characterized by power law distributions of
tags, over which commonly used similarity metrics, including the Jaccard coefficient and
the cosine similarity, fail to compute. We thus propose a novel metric, specifically developed to
capture similarity in large-scale folksonomies, that is based on a mutual reinforcement
principle: that is,  two tags are deemed similar if they have been associated to similar resources,
and vice-versa two resources are deemed similar if they have been labelled by similar tags. We offer
an efficient realisation of this similarity metric, and assess its quality experimentally, by
comparing it against cosine similarity, on three large-scale datasets, namely Bibsonomy, MovieLens
and CiteULike.
\end{abstract}

\section{Introduction} \label{sec:Introduction}

% intro to folksonomy
The rise of Web 2.0 has transformed users from passive consumers to active producers of content.
This has exponentially increased the amount of information that is available to users, from videos
on sites like YouTube and MySpace, to pictures on Flickr, music on Last.fm, blogs on Blogger, and
so on. This content is no longer categorised according to pre-defined taxonomies (or ontologies).
Rather, a new trend called {\em social} (or {\em folksonomic}) {\em tagging} has emerged, and
quickly become the most popular way to describe content within Web 2.0 websites. Unlike taxonomies,
which overimpose a hierarchical categorisation of content, folksonomies empower end users by
enabling them to freely create and choose the tags that best describe a piece of information (a
picture, a blog entry, a video clip, etc.). However, this freedom comes at a cost: since tags are
informally defined, continually changing, and ungoverned, finding content of interest has become a
main challenge, because of the number of synonyms, homonyms, polysemy, as well as the inevitable
heterogeneity of users and the noise they introduce.

% hot to find content - need of similarity computation
In order to assist users finding content of their own interest within this information abundance,
new techniques, inspired by traditional recommender systems, have been developed: for example,
whenever a user searches from some content using query tags $\{t_1, \dots, t_m\}$, new tags
$\{t_{m+1}, \dots, t_{m+n}\}$ are being added to the query, based on their {\em similarity} to
their original query tags. This is done to increase the chances of finding content of relevance in
these extremely sparse settings. Various metrics have been used to compute the similarity among
folksonomy entities, including, for instance, cosine similarity, Jaccard coefficient, and Pearson
Correlation. Performance results demonstrate an increase in accuracy and coverage of searches when
using these techniques; however, evaluation has been conducted on manipulated datasets
%, obtained by computing a $p$-core (see \cite{gemmell*09}) over the original folksonomy, 
so to obtain a much denser one.  
%($p=5$ in most cases). 
We argue that such manipulations alter the nature of real folksonomies, and indeed
eliminate the problem, rather than solving it.

% intrinsic problems
Unmodified  real-world folksonomies are characterized by two key properties: the {\em power law}
distribution of tags, and the {\em non-independence} of data. Empirical studies
\cite{cattuto2007network,Ursino-InfSys-09} illustrate that tag usage in folksonomies follows a
power law distribution; this means that, if we were select any two tags, the probability that the
resources jointly labelled by them is non-zero is extremely low. As a result, computing tag
similarity on un-modified folksonomies, using traditional metrics like cosine similarity, would
almost always yield close-to-zero values, thus failing to support users in retrieving resources
relevant to their queries. Furthermore, metrics like cosine assume that tags are semantically
independent of each other; once again, this assumption does not hold in  real folksonomies, where
tags may be synonyms to each other.

% proposition
In this paper, we propose a novel similarity metric that can be used to accurately quantify tag
similarity in large-scale real-world folksonomies (Section~\ref{sec:approach}). This similarity
metric is computed following an  iterative algorithm, grounded on a {\em mutual reinforcement}
principle: that is,  two tags are similar if they label similar resources,  and vice-versa,  two
resources are similar if they have been labelled by similar tags. We describe  an efficient
realisation of this similarity metric (Section~\ref{sub:taguser}), and empirically quantify its
quick convergence on three large-scale datasets, namely
BibSonomy\footnote{{\tt http://www.bibsonomy.org/}}, MovieLens\footnote{{\tt http://www.movielens.org/}}, and
CiteULike\footnote{{\tt http://www.citeulike.org/}}. We measure Precision and Recall of our metric, and
compare it to cosine similarity on these unprocessed datasets
%, on different $p$-cores of the above datasets 
(Section \ref{sec:experiments}). 
Our findings demonstrate that,
%: on $p$-core of 5 (i.e., highly manipulated, dense datasets),  the two metrics yields similarly-good results; however, 
when considering our un-manipulated datasets,  the performance of our novel
similarity metric provides higher Precision and Recall w.r.t. the cosine similarity. Section~\ref{sec:RelatedWork}
covers related works on similarity measures, mainly  applied to folksonomies. Finally, in
Section~\ref{sec:Conclusions} we draw our conclusions.

\section{Background} \label{sec:background}
In this section,  we formally introduce some concepts that will be extensively used in the following, when presenting our approach.
The first concept we consider is that of a folksonomy \cite{hotho2006information}:

\begin{definition}
	\label{def:folk} Let $US = \{u_1, \ldots, u_{n_u} \}$ be a set of users, let $RS = \{r_1,\ldots, r_{n_r}\}$
	be a set of resource URIs and let $TS = \{t_1, \ldots, t_{n_t}\}$ be a set of tags. A {\em folksonomy}
	$F$ is a tuple $F = \langle US,RS,TS, AS \rangle $, where $AS \subseteq US \times RS \times TS$ is
	a ternary relationship called {\em tag assignment set}.
\end{definition}

In this definition we do not make any assumption about the nature of resources; they could be
URLs (like in Delicious), photos (as in Flickr), music files (as in Last.fm),  documents (as in CiteULike), and so on.

According to Definition~\ref{def:folk}, a folksonomy $F$ is a ``three-dimensional'' data structure
whose ``dimensions'' are represented by users, tags and resources. In particular, an element $a \in
AS$ is a triple $\langle u, r, t \rangle$, indicating that user $u$ labelled resource $r$ with tag
$t$. To simplify modeling and management of folksonomies, their inherent tripartite graph structure
is often mapped into three {\em matrices}, whereby each matrix models one relationship at a time
(i.e., between tags and resources, tags and users,  and resources and users)
\cite{mika2007ontologies}. In this paper, we adopt the same matrix-based representation.
Specifically, being $n_r$, $n_t$ and $n_u$ the number of resources, tags and users respectively, we
represent a folksonomy as the following three matrices:

\begin{itemize}
	\item $\mathbf{TR}$ (Tag-Resource): a  $n_t \times n_r$ matrix such that $\mathbf{TR}_{ij}$
is the number of times the tag $i$ labelled resource $j$;
	\item $\mathbf{TU}$ (Tag-User): a $n_t \times n_u$ matrix such that $\mathbf{TU}_{ij}$ is the number of times the tag $i$ has been used by  user $j$;
	\item $\mathbf{RU}$ (Resource-User): a $n_r \times n_u$ matrix such that $\mathbf{RU}_{ij}$
is the number of times resource $i$ has been labelled by the user $j$.
\end{itemize}

Tag similarity within a folksonomy can then be computed by looking at the resources these tags have
been attached to. In particular, each tag $t_i$ can be mapped onto a vector $\mathbf{t}_{r}(i)$
corresponding to the $i$-th row of $\mathbf{TR}$. Given an arbitrary pair of tags $t_i$ and $t_j$,
their  similarity $s(t_i,t_j)$ can be computed as the {\em cosine similarity} (CS) of the vectors
$\mathbf{t}_{r}(i)$ and $\mathbf{t}_{r}(j)$:
\begin{equation}
	\label{eqn:ressim}
  s(t_i,t_j) = \frac{\langle \mathbf{t}_{r}(i) , \mathbf{t}_{r}(j)\rangle}{\sqrt{\langle \mathbf{t}_{r}(i) , \mathbf{t}_{r}(i)\rangle} \sqrt{\langle \mathbf{t}_{r}(j) , \mathbf{t}_{r}(j)\rangle}}
\end{equation}
being $\langle \cdot, \cdot \rangle$ the usual {\em inner product} in $\mathbb{R}^{n_r}$.

Cosine similarity has been successfully applied in the context of Information Retrieval
\cite{MaRaSc08}. Within a folksonomy, Equation~\ref{eqn:ressim} states that the similarity score of
a pair of tags is high if they {\em jointly co-occur} in labelling the same subset of resources.
However, two key properties of folksonomies, that are, {\em (i)} the power law distribution of tags and
{\em (ii)} their non-independence, cause Equation~\ref{eqn:ressim} to yield very poor results in this
domain, as we shall discuss next.

{\bf Power Law in Tag Distribution.}
Let us consider a real-world folksonomy like BibSonomy. BibSonomy
\cite{hotho2006bibsonomy,jaschke2007analysis} is a social bookmarking service in which users are
allowed to tag both URLs and scientific papers. A power law distribution of tags on scientific references emerges. In particular, resources were described by no more than 5 different tags (roughly 81\%), and usually less than 3 (roughly 58\%). A small portion of frequently adopted tags used to bookmark scientific references, and a long tail of tags (roughly 81\%) being used less than 5 times.

Following the above observations, matrix $\mathbf{TR}$ is rather sparse; thus, if we were to select
any pair of tags $t_i$ and $t_j$, most of the components of the corresponding vectors
$\mathbf{t}_{r}(i)$ and $\mathbf{t}_{r}(j)$ would be 0 and, therefore their inner product would be
close to 0. The cosine similarity between {\em any} $t_i$ and $t_j$ would therefore be almost 0,
regardless of the initial choice of $t_i$ and $t_j$. Such  counter-intuitive result is an effect of
the inadequacy of  cosine similarity to capture properties of tags in large-scale real
folksonomies.

{\bf Non-Independence of Tags.} Cosine similarity implicitly assumes that the components of the
vectors appearing in Equation~\ref{eqn:ressim} are {\em independent} of each other. Such an
assumption does not often hold true. For instance, consider a  folksonomy consisting of two
resources $r_1$ and $r_2$, representing two different scientific papers, both discussing about
folksonomies. Suppose that the paper associated with $r_2$ is an extension of the paper associated
with $r_1$. Finally, assume to bookmark the resource $r_1$ with the tag $t_1 = $ ``folksonomy'' and
to bookmark the resource $r_2$ with the tag $t_2 = $ ``social tagging''. In this case, the
similarity between $t_1$ and $t_2$ computed according to Equation~\ref{eqn:ressim} would be 0, even
if $t_1$ and $t_2$ should result similar each other. The mutual similarity between $t_1$ and $t_2$
can be assessed only if we consider the non-independence of the resources they label.

\section{Approach Description} \label{sec:approach}
In this section, we present  a new definition of tag (and resource) similarity, that is
particularly suited to quantify similarity of elements (be them tags of resources) in datasets
characterized by power law distribution and non-indepen\-dence of data. Our definition of similarity
relies on the {\em mutual reinforcement principle}:

\begin{quote}
	{\em Two tags are similar if they label similar resources, and conversely, two resources are
similar if they are labelled by similar tags.}
\end{quote}

In the following, we shall derive a mathematical formula to compute tag and resource similarity on
the basis of the principle stated above. After this, we shall illustrate why our formula is able to
effectively address the power law and non-independence challenges.

We designed an {\em iterative algorithm} to compute the similarity score. In the base case, given a
pair of tags $\langle t_a, t_b \rangle$ and a pair of resources $\langle r_a, r_b \rangle$, we
define the {\em tag similarity} ${st}^0(t_a,t_b)$ and the {\em resources similarity}
${sr}^0(r_a,r_b)$ as follows:
    \begin{equation}
    	\label{eqn:basecase}
    	{st}^0(t_a,t_b) = \delta_{ab} \quad \quad {sr}^0(r_a,r_b) = \delta_{ab}
    \end{equation}
being $\delta_{ab}$ the {\em Kronecker symbol}\footnote{We recall that the Kronecker symbol
$\delta_{ab}$ is equal to 1 if $a$ and $b$ coincide and 0 otherwise.}. Equation~\ref{eqn:basecase}
reflects the fact that, in the initial step, each tag (resp., resource) is similar only to itself
and it is dissimilar to all other tags (resp., resources).

At the $k$-th step, let ${st}^{k-1}(t_a,t_b)$ (resp., ${sr}^{k-1}(r_a,r_b)$) be the tag (resp.,
resource) similarity between the tags $t_a$ and $t_b$ (resp., resources $r_a$ and $r_b$). We apply
the following rules to update ${st}^{k-1}(t_a,t_b)$ (resp., ${sr}^{k-1}(r_a,r_b)$):
    \begin{eqnarray}
    	{st}^{k}(t_a,t_b) = \frac{ ST^k(t_a,t_b) }{ \sqrt{ST^k(t_a,t_a)} * \sqrt{ST^k(t_b,t_b)} } && \label{eqn:coSim1}\\
    	{sr}^{k}(r_a,r_b) = \frac{ SR^k(r_a,r_b) }{ \sqrt{SR^k(r_a,r_a)} * \sqrt{SR^k(r_b,r_b)} } && \label{eqn:coSim2}
    \end{eqnarray}
where:
    \begin{eqnarray}
        ST^{k}(t_a,t_b) = \sum_{i,j=1}^{n_r}{ \mathbf{TR}_{ai} * \Psi_{ij} * {sr}^{k-1}(r_i,r_j) * \mathbf{TR}_{bj} } && \label{eqn:defSTSR1} \\
        SR^{k}(r_a,r_b) = \sum_{i,j=1}^{n_t}{ \mathbf{TR}_{ia} * \Psi_{ij} * {st}^{k-1}(t_i,t_j) * \mathbf{TR}_{jb} } && \label{eqn:defSTSR2}
    \end{eqnarray}

Here $\Psi_{ij}$ is equal to $1$ if $i = j$ and it is equal to $\psi$ if $i \neq j$, where $\psi$ (called {\em propagation factor})
is a value belonging to the  interval $[0,1] \in \mathbb{R}$.

Equations~\ref{eqn:coSim1}--\ref{eqn:coSim2} rely on the following intuitions. Given a pair of tags $\langle t_a, t_b \rangle$, at the $k$ iteration, we consider {\em all pair of resources} $\langle r_i, r_j\rangle$ in our folksonomy and we take their similarity ${sr}^{k-1}(r_i,r_j)$ into account to compute ${st}^{k}(t_a,t_b)$. In particular, we compute a {\em weighted sum} of all the similarity values ${sr}^{k-1}(r_i,r_j)$, where the weights reflect the {\em strength of the association} between the tag $t_a$ and the resource $r_i$, and the tag $t_b$ and the resource $r_j$. As a consequence, the higher the similarity between $r_i$ and $r_j$, the higher the contribution of the association between the tag $t_a$ and the resource $r_i$, and the tag $t_b$ and the resource $r_j$. Finally, the term $\Psi_{ij}$ is instrumental to give higher relevance to tags that labelled the very {\em same} resources, w.r.t. the fact that they labelled two  {\em similar} (but different) resources.

Note that, in the special case in which $\psi = 0$, our method does not depend on $k$ and
Equations~\ref{eqn:coSim1}--\ref{eqn:coSim2} reduce to the cosine similarity formulation. In fact,
in this particular case, all the contributions $sr^{k-1}(r_i, r_j)$ and $st^{k-1}(r_i, r_j)$ are
disregarded when $i \neq j$, and are taken into consideration only when $i = j$. Since all
contributions $sr^{k-1}(r_i, r_i)$ and $st^{k-1}(r_i, r_i)$ are equal to 1 by definition, it
follows that Equations~\ref{eqn:coSim1}--\ref{eqn:coSim2} reduce to the cosine similarity formulation.

Equations~\ref{eqn:coSim1}--\ref{eqn:coSim2} are able to effectively address the power law and
non-independence of data challenges we outlined above. In fact:

\begin{itemize}
	\item In the computation of tag (resp., resource) similarity, we leverage on the similarity
of all pairs of resource (resp., tag) similarities.
	As a consequence, unlike cosine similarity, we do not restrict ourselves to consider only the resources
jointly labelled by two tags (resp., the tags jointly labelling two resources), which can be few,
but we {\em iteratively propagate} similarity scores by considering all the pairs of similar resources jointly labelled by the two tags (resp., all the pairs of similar tags jointly labelling two resources).
%including resources which are labelled by a large number of tags (resp., tags labelling a large number of resources).
In this way we
are able to face the power law occurring in tag usage.

	\item In our definition of similarity, if two tags label similar,
{\em even if not coincident}, resources their similarity score will be greater than 0, whereas
the cosine similarity would return 0. As a consequence, our similarity method takes into account
forms of correlation among pairs of resources and/or tags rather than assuming their
independence.
\end{itemize}

%The same assumption on the similarity among tags and resources depicted above can be extended to
%the cases: {\em (i)} resources vs. users, and, {\em (ii)} users vs. tags.

\section{Realization} \label{sub:taguser}

From a computational standpoint, Equations~\ref{eqn:coSim1}--\ref{eqn:coSim2} could entail a large
overhead for two reasons:

\begin{itemize}
	\item From a theoretical standpoint, our approach may need an infinite number of iterations.
	As a consequence, we need a stopping criterion allowing us to safely terminate the
execution of Equations~\ref{eqn:coSim1}--\ref{eqn:coSim2} after a finite (and low) number of
iterations.

	\item Equation~\ref{eqn:coSim1} (resp., Equation~\ref{eqn:coSim2}) requires the computation
of $n_r^2$ resource-resource (resp,.  $n_t^2$ tag-tag) similarities, at each  $k$-th step.
	This could make our similarity measure inapplicable in practical cases, because each
iteration requires exactly $n_r^2 \times n_t^2$ computations.
\end{itemize}

Fortunately, there are two important results making our similarity measure applicable and entailing the same complexity level as cosine similarity.   The first result can be stated by the theorem showed and proved in the Appendix\footnote{See
{\tt http://tinyurl.com/proof-seke2011}.} which affirms that the sequences ${st}^{k}(t_a,t_b)$ and ${sr}^{k}(r_a,r_b)$ defined as in Equations~\ref{eqn:coSim1}--\ref{eqn:coSim2} converge.

This theorem ensures that, after a certain number of iterations, Equations
\ref{eqn:coSim1}--\ref{eqn:coSim2} converge to stable values. During experimentation conducted on
three real folksonomies (see Section~\ref{sub:dataset}), we empirically found that convergence was
achieved after as little as five iterations, thus suggesting that our similarity measure is
applicable in practical cases.

Furthermore, Equations~\ref{eqn:coSim1}--\ref{eqn:coSim2} can be defined, without any loss of
generality, as a simple matrix product (such as in cosine similarity). Specifically, let
$\mathbf{st}^k$ and $\mathbf{sr}^k$ be the tag-tag and resource-resource similarity matrices
respectively, with $\mathbf{st}^0=\mathbf{I}_t$ and $\mathbf{sr}^0=\mathbf{I}_r$; here
$\mathbf{st}^0=\mathbf{I}_t$ (resp., $\mathbf{sr}^0=\mathbf{I}_r$) is the $n_t \times n_t$ (resp.,
$n_r \times n_r$) {\em identity matrix}. If we indicate with the symbol ``$\circ$'' the {\em
Hadamard matrix product}\footnote{Given two matrices $\mathbf{A}$ and $\mathbf{B}$ of the same
dimensions, the {\em Hadamard product} $\mathbf{A} \circ \mathbf{B}$ is a matrix of the same
dimensions of $\mathbf{A}$ and $\mathbf{B}$ and it is defined as follows: $(\mathbf{A} \circ
\mathbf{B})_{ij} = \mathbf{A}_{ij} \cdot \mathbf{B}_{ij}$} \cite{GoVa96}, at the $k$-th step, the
$\mathbf{st}^k$ and $\mathbf{sr}^k$ matrices can be computed as:
    \begin{eqnarray}
    	& \mathbf{{st}}^{k} = \mathbf{ST}^k \circ \mathbf{DT}^k & \label{eqn:matrix1}\\
    	& \mathbf{{sr}}^{k} = \mathbf{SR}^k \circ \mathbf{DR}^k & \label{eqn:matrix2}
    \end{eqnarray}
where:
    \begin{eqnarray}
    	& \mathbf{ST}^{k} =\mathbf{TR} \times \left(\mathbf{\Psi_r} \circ \mathbf{{sr}}^{k-1}\right) \times \mathbf{TR}^t & \label{eqn:ST}\\
    	& \mathbf{SR}^{k} = \mathbf{TR}^t \times \left(\mathbf{\Psi_t} \circ \mathbf{{st}}^{k-1}\right) \times \mathbf{TR} & \label{eqn:SR}\\
	& \mathbf{DT}^k_{ab} = \frac{1}{\sqrt{\mathbf{ST}^{k}_{aa}}\sqrt{\mathbf{ST}^{k}_{bb}}} &\\
	& \mathbf{DR}^k_{ab} = \frac{1}{\sqrt{\mathbf{SR}^{k}_{aa}}\sqrt{\mathbf{SR}^{k}_{bb}}} &
    \end{eqnarray}

In the above equations, we have indicated with $\mathbf{\Psi}_r$ (resp., $\mathbf{\Psi}_t$) a
square matrix $n_r \times n_r$ (resp., $n_t \times n_t$) where all the elements are set to $\psi$,
with the exception of the diagonal where the elements are set to 1; the symbol $\mathbf{TR}^t$
represents the transpose of matrix $\mathbf{TR}$. We have thus reduced the computational complexity
of each iterative step from $n_r^2 \times n_t^2$  to  a simple matrix product; this reduction,
coupled with the empirical observation that 5 iterative steps are sufficient to find convergence,
makes our  similarity metrics suitable in practical contexts. The last question that needs
answering is how effective (in terms of Precision and Recall) our similarity metric is w.r.t.
traditional ones like cosine. We answer this question next.

\section{Experiments} \label{sec:experiments}
In order to evaluate the performance of our similarity measure, we built a prototype in Java and
MySQL and we conducted experiments using three well known social tagging websites: Bibsonomy,
CiteULike, and MovieLens. The experiments we carried out aimed to answer the following question:

\begin{quote}
 {\em If we consider any two tags $t_i$ and $t_j$ belonging to a folksonomy, is our
similarity measure capable of accurately assessing the extent to which they are related (similar) each other? And can it do so
even when such tags have been drawn from the long tail of low
popularity tags?}
\end{quote}

\subsection{The Dataset} \label{sub:dataset}
To answer the above question, we conducted experiments on the following three datasets.

{\bf Bibsonomy.} Bibsonomy is a social bookmarking website promoting the sharing of both scientific
reference and general URL. We downloaded a snapshot of the website in June 2009, containing
bookmarks made between January 1999 and June 2009.

{\bf CiteULike.}
CiteULike is a social bookmarking website that aims to promote and develop the sharing of scientific references amongst researchers. CiteULike enables scientists to organize their libraries with freely chosen tags which produce a folksonomy of academic interests. CiteULike runs a daily process which produces a snapshot summary of what articles have been posted by whom and with what tags up to that day. We downloaded one such archive in November 2009, containing bookmarks made between November 2004 to November 2009.

{\bf MovieLens.} MovieLens is a rate-based recommendation website that suggests to users movies
they might like. We downloaded such dataset in January 2009, containing
bookmarks made from December 2005 to January 2009. 

Table~\ref{tab:Datasets} summarizes the features of the involved datasets.

\begin{table}[t]
    \centering
    \footnotesize
    \begin{tabular}{||c|c|c|c|c||}
    \hline \hline
    {\em Dataset}    & {\em Users} & {\em Resources} & {\em Tags} & {\em Bookmarks}\\
    \hline \hline
    Bibsonomy        & 4,696       & 578,587        & 147,076    & 648,924\\
    \hline
    CiteULike        & 57,053      & 1,928,302       & 401,620    & 2,281,609\\
    \hline
    MovieLens        & 4,009       & 7,601           & 15,240     & 55,484\\
    \hline \hline
    \end{tabular}
    \caption{Features of our datasets}
    \label{tab:Datasets}
\end{table}

\subsection{Simulation Setup} \label{sub:simulation}

Our experimental investigation aimed to quantify, in each of the above datasets, the extent to
which our similarity measure was capable of identifying related tags, especially when tags were
drawn from  the long tail. 
To investigate this, for each dataset of Table~\ref{tab:Datasets} has been used as follows. We split it into two different sets, called {\em test set} and {\em train set}. Each train set was composed of 90\% random
bookmarks taken from the involved dataset; we used these bookmarks for  training purposes. Test
sets contained the remaining 10\% of bookmarks which were used for testing. Each bookmark in a test
set has then been used as a query; specifically, if the number of tags in such bookmark was large
enough, then these were split into two different sets -- if possible of the same size -- called
$tSet_Q$ ({\em query tag set}) and $tSet_E$ ({\em expected tag set}). In our experiments, a
bookmark was considered large enough if it had at least 3 tags associated. Tags composing $tSet_Q$
were used to query the train set; in particular, we selected from the train set the $k$ tags most
similar to tags belonging to $tSet_Q$, according to two metrics: the one we proposed in Section
\ref{sec:approach}, and cosine similarity, which we used as benchmark. We denote this set as
$tSet_R$ ({\em result tag set}). The value of $k$ was chosen equal to the size of the expected set
in such a way that $tSet_R$ and $tSet_E$ had the same size. Finally, we compared $tSet_R$ with
$tSet_E$: the higher the overlap between $tSet_R$ and $tSet_E$, the more effective the similarity
measure in identifying related tags. This follows the intuition that, if a user associated a set of
tags to a certain resource, such tags are related to each other (that is, $tSet_E$ contains tags
related to those contained in $tSet_Q$).

To quantitatively evaluate our similarity measure, we computed two metrics commonly used in
Information Retrieval,  namely Precision and Recall \cite{BaRi99}:
    \begin{eqnarray}
    	& Precision = \frac{|tSet_R \cap tSet_E|}{|tSet_R|} &\\
    	& Recall = \frac{|tSet_R \cap tSet_E|}{|tSet_E|} &
    \end{eqnarray}

We computed Precision and Recall values for each test bookmark;  we repeated this process 10
times over different train and test splits of the datasets. The results we present next are averages of such runs.

\subsection{Results} \label{sub:results}

Tables~\ref{tag:precision} and~\ref{tag:recall} shows values of Precision and Recall we obtained by applying our similarity measure on the datasets of Table~\ref{tab:Datasets}, for different values of $\psi$ (see
Equations~\ref{eqn:coSim1}--\ref{eqn:coSim2}). The benchmark is our similarity measure with
$\psi = 0$, that is, the case in which our similarity measure reduces into cosine
similarity. 

\begin{table}[t]
    \centering
    \footnotesize
    \begin{tabular}{||c|c|c|c||}
    \hline \hline
    {\em Propagation}    & {\em Bibsonomy} & {\em CiteULike} & {\em MovieLens}\\
    \hline \hline
    $\psi = 0$           & 0.100638896     & 0.057922233     &  0.075126961 \\
    \hline
    $\psi = 0.15$        & 0.128318833     & 0.063290603     &  0.112358995 \\
    \hline
    $\psi = 0.3$         & 0.139761842     & 0.070652236     &  0.115026291 \\
    \hline
    $\psi = 0.6$         & 0.140748308     & 0.079320913     &  0.115534133 \\
    \hline \hline
    \end{tabular}
    \caption{Precision values in our datasets}
    \label{tag:precision}
\end{table}

\begin{table}[t]
    \centering
    \footnotesize
    \begin{tabular}{||c|c|c|c||}
    \hline \hline
    {\em Propagation}    & {\em Bibsonomy} & {\em CiteULike} & {\em MovieLens}\\
    \hline \hline
    $\psi = 0$           & 0.100625714     & 0.057864697     &  0.075143054 \\
    \hline
    $\psi = 0.15$        & 0.128373044     & 0.063273342     &  0.110927464 \\
    \hline
    $\psi = 0.3$         & 0.139939546     & 0.070634975     &  0.119373901 \\
    \hline
    $\psi = 0.6$         & 0.140429576     & 0.079303652     &  0.119949092 \\
    \hline \hline
    \end{tabular}
    \caption{Recall values in our datasets}
    \label{tag:recall}
\end{table}

From the analysis of Tables~\ref{tag:precision} and~\ref{tag:recall} we can draw the following main observation: in large scale folksonomies, classical approaches -- such as cosine similarity ($\psi = 0$) -- have difficulties finding similarity relationships among the tags belonging to the long tail, as
their Precision and Recall is lower than those achieved with our iterative approach for any
value of $\psi$. The considered datasets are characterized by a very long and prominent tail of low popularity tags; in these real cases, out iterative measure of similarity produces Precision/Recall that
is approximately 40\% better than cosine similarity for BibSonomy and CiteULike, and
approximately 50\% better for MovieLens.

\section{Related Work} \label{sec:RelatedWork}

In the last few years, folksonomies have been the subject of extensive research. An interesting
survey on the characteristics of folksonomies can be found in \cite{cattuto2007network}. One of the
first investigations into the characteristics of folksonomies has been presented by Mathes
\cite{mathes2004folksonomies}: in that work, the author discusses advantages (e.g., simplicity of
use) and disadvantages (e.g., ambiguity, synonyms) of folksonomies, and investigates the community
aspects behind folksonomies,
% (e.g., underlying motivations explaining its success, how  cooperation works, how sharing thoughts reflects on users), 
on two  scenarios, Flickr\footnote{{\tt http://www.flickr.com/}} and Delicious\footnote{{\tt http://www.delicious.com/}}.

Despite their easy-of-use, the lack of structure that characterises folksonomies makes it difficult
to browse and find relevant content. To tackle this issue, the research community has been actively
researching techniques to support information retrieval. Approaches in this area have followed one
of two streams: they have either tried to empirically derive an ontology from the underlying
folksonomy, or they have tried to apply graph-exploration techniques on the folksonomy itself.

Lambiotte \cite{lambiotte2006collaborative} and Mika \cite{mika2007ontologies}, for example, were
the first to extend the classic bipartite model of tag-resource towards a tripartite model, which
takes into account both users (as actors), tags (as concepts) and resources (as instances); they
showed that, by  applying this model to Delicious,  a lightweight ontology could be extracted from
the underlying folksonomy. Similarly, \cite{heymann2006collaborative} used similarity metrics to
reconstruct a concept hierarchy.

Hotho et al. \cite{hotho2006information,schmitz2006mining} followed a different approach instead:
they presented a formal model, which converts a folksonomy into an undirected weighted graph, and
coupled it with a new search algorithm, namely ``FolkRank'', based on the well-known seminal
``PageRank'' \cite{brin1998anatomy}. They applied this algorithm to Delicious, and showed how it
can be  used as a tag recommender system. Other extensions of recommender systems to folksonomy
structures have been explored  \cite{xu2006towards,heymann2008social}; some of
these have been assessed against one of the datasets we adopted in this study, namely BibSonomy
\cite{hotho2006bibsonomy,jaschke2007analysis}.

%Cattuto et al. \cite{cattuto2008emergent} introduced the concept of ``resource distance'', based on the collective tagging activity of users, analyzing Delicious.
%This way, they successfully depicted the community structure of that resource network.

All the above approaches rely on a similarity measure to quantify tag relatedness. Measures which
have been often used in the literature include  the Jaccard coefficient \cite{hassan2006improving},
the cosine similarity \cite{diederich2006finding}, and a number of improvements over it
\cite{liu2004learning,zhang2009method}. Liu et al. \cite{liu2004learning} dwelt  further into the
problem of computing similarities in folksonomies; in particular, they questioned the common
assumption that  text categorization can be mapped onto orthogonal spaces, due to  problems of
synonyms and ambiguities (as already figured out by \cite{mathes2004folksonomies}). They then
devised an improved similarity metric (``SNOS'', Similarity equations in the Non-Orthogonal Space)
which is optimized for comparing objects mapped onto non-orthogonal spaces, considering a principle
of ``mutual reinforcement'' from which we drew inspiration in this work. They proved the
convergence of this technique and experimentally investigated the performance of SNOS on synthetic
datasets, such as the formerly called MSN search engine (now, Bing\footnote{{\tt http://www.bing.com/}}).
Their novel metric was shown to outperform the classic cosine similarity, if applied to the context
of finding similar queries. Some of their findings are here extended to the domain of folksonomies.

Similarity measures have often been evaluated on different datasets, making it difficult to assess
their relative advantages and disadvantage in different domains. Furthermore, they have often been
applied to manipulated datasets, making the comparison even more difficult. Indeed, in order to
critically compare them,  an evaluation framework has recently been proposed
\cite{markines2009evaluating}, with the aim of providing support to systematically  compare several
tag similarity measures, using data from Delicious \cite{cattuto2008semantic}. This work
contributes to the assessment of the suitability of similarity measures to scenarios characterized
by  power-law distribution of tags and non-independence of data, showing how traditional measures
like cosine do not work, and proposing an alternative, iterative measure that provides good
accuracy instead.

% Licia: I cut, as i don't understand how they relate to the rest. If you put them back, you need to glue them to the rest
%Semiotic dynamics of the tagging vocabulary have been already investigated \cite{cattuto2007semiotic}.
%These findings reflect in complex applications, such as the Watson DeepQA Project \cite{ferrucci2010building}, where the most approprieate similarity measure must be adopted in order to solve problems of Q\&A, in real-time and in a fast and reliable way.

\section{Conclusions} \label{sec:Conclusions}

In this paper, we have shown that  real world folksonomies are characterized by power law
distributions of tags and non-independence of data. Under these conditions, traditional  similarity
measures like  cosine similarity fail to capture tags relatedness. To remedy this, we have
proposed a novel metric, specifically developed to capture similarity in large-scale folksonomies,
that is based on the mutual reinforcement principle: that is, two tags are deemed similar if they
have been associated to similar resources, and vice-versa two resources are deemed similar if they
have been labelled by similar tags. We have described an efficient realisation of this similarity
metric, and assessed its quality experimentally, by comparing it against cosine similarity, on
three large-scale datasets, namely Bibsonomy, MovieLens and CiteULike.\\

{\bf Acknowledgement.} The research leading to these results has received funding from the European Community's Marie Curie Fellowship Programme (FP7-PEOPLE-2009-IEF) under the Grant Agreement n. 38675. The authors are solely responsible for it and it does not represent the opinion of the Community. The Community is not responsible for any use that might be made of information contained therein.

\bibliography{sigproc}
\bibliographystyle{plain}
%%%\bibliographystyle{latex8}

%\begin{figure*}[!ht]%
%	% \includegraphics[trim=20mm 25mm 25mm 20mm, clip,  angle=90]{fig/bibsonomy-distribution.pdf}% FULL PAGE VERTICAL
%%	\includegraphics[trim=20mm 25mm 25mm 20mm, clip, width = 2\columnwidth]{fig/bibsonomy-distribution.pdf}%
%\includegraphics[trim=20mm 25mm 25mm 20mm, clip, width = 2\columnwidth]{./fig/bibsonomy-distribution.png}
%% HALF PAGE ORIZONTAL
%	\caption{Count of tags appearing with high frequency}%
%	\label{fig:bibsonomy-distribution}%
%\end{figure*}

\end{document}